\documentclass[a4paper,aps,pra,preprintnumbers]{revtex4} %добавить showpacs, чтобы показывать pacs
\usepackage{graphics,graphicx,epsfig}	% для рисунков
\usepackage[centertags]{amsmath}
\usepackage{amsfonts}
\usepackage{amssymb, dsfont}			% для математических формул
\usepackage[cp1251]{inputenc}			% для восприятия кодировки Windows
\usepackage[english,russian]{babel}		% для поддержки переносов в выбранном языке
\usepackage{graphicx}
\usepackage{amsthm,color}
\usepackage{ulem}					% ulem - для зачеркивания
\usepackage{textcomp}                           	% всякие символы
\usepackage{euscript}
\usepackage{hyperref}				% для гиперссылок
\hypersetup{colorlinks=true, citecolor=blue, linkcolor=black} % цвет ссылок c помощью \cite и с помощью \ref

\begin{document}
\def\BE{\begin{equation}}
\def\EE{\end{equation}}
\def\BEA{\begin{eqnarray}}
\def\EEA{\end{eqnarray}}
\def\BY{\begin{eqnarray}}
\def\EY{\end{eqnarray}}

\def\L{\label}
\def\nn{\nonumber}
\def\ds{\displaystyle}

\def\({\left (}
\def\){\right )}
\def\[{\left [}
\def\]{\right]}
\def\<{\langle}
\def\>{\rangle}

\def\td{\tilde}
\def\pr{\prime}
\def\2!{\!\!}
\def\3!{\!\!\!}
\def\4!{\!\!\!\!}
\def\5!{\!\!\!\!\!}
\def\6!{\!\!\!\!\!\!}

\def\k{\mathbf{k}}
\def\q{\mathbf{q}}
\def\r{\vec{r}}
\def\ro{\vec{\rho}}
\def\h{\hat}
\def\hs{\hat{\sigma}}
\def\a{\hat{a}}
\def\b{\hat{b}}
\def\c{\hat{c}}
\def\e{\hat{e}}
\def\v{\hat{v}}
\def\pr{\prime}
\def\Om{\Omega}

%
%-----------------------------------------------------------
\title{The Tripod High-Speed Quantum Memory\\
As a Beam Splitter with Arbitrary Splitting Ratios} \vspace{1cm}
\author{A. S. Losev}
\address{Saint Petersburg State University, 7/9 Universitetskaya nab., St. Petersburg, 199034 Russia}
\address{Saint Petersburg State Marine Technical University, 3 Lotsmanskaya str., St. Petersburg, 190121 Russia}
\date{\today}
%
%
%-----------------------------------------------------------
\begin{abstract}
The utilisation of a quantum memory cell as a beam splitter with arbitrary coefficients is demonstrated theoretically. For such a beam splitter, an input-output matrix is derived. We investigate the high-speed quantum memory based on the tripod-type atomic levels and transitions.
\end{abstract}
%\pacs{42.50.Dv, 42.50.Gy, 42.50.Ct, 32.80.Qk, 03.67.-a}
\maketitle
%
%-------------------------------------------
\section{Introduction}\L{I}
Over the past two decades, considerable  attention has been paid to the realisation of light beam routers on atomic media in the tripod configuration. Such media allow a signal pulse to be split into two, separated in time \cite{Paspalakis-2002, Paspalakis-2002PRA, Wang-2004, Unanyan-2004, Raczynski-2006, Moller-2007, Yang-2015, Shou-2019PRA99, Shou-2019PRA100} and possessing new quantum-statistical properties \cite{Wang-2004, Yang-2015}. Moreover, it has been proposed that memory cells, in the electromagnetically induced transparency regime and with a tripod-type atomic configuration, should be utilised as a conventional beam splitter, with two pulses at the input and two at the output \cite{Raczynski-2007, Raczynski-2007PRA}. In \cite{Losev-2016, Losev-2017, Losev-2020}, we have also demonstrated that the tripod memory can be used as a 50/50 beam splitter by applying the high-speed quantum memory protocol.

In this Letter, in continuation of the works \cite{Losev-2017, Losev-2020}, we also consider the tripod high-speed quantum memory cell as a beam splitter, with arbitrary transmission and reflection coefficients. We show how the input-output matrix is parameterised through the Rabi frequencies of the driving fields.
%
%
%-------------------------------------------
\section{Mapping and readout procedures}\L{II}
We consider the interaction of a cold atomic ensemble in the one-dimensional, paraxial approximation with the signal and driving pulses of the electromagnetic field. The atoms are regarded as quasi-immobile, representing four-level systems of tripod configuration (see Fig. \ref{Fig1}). The interaction is considered to be resonant.

The following description will outline the procedure of mapping (MP) and readout (RP). At the point of initiation of the first mapping process, all atoms are in state $|3\>$.
\begin{figure}[h]
\centering
\includegraphics[height=34mm]{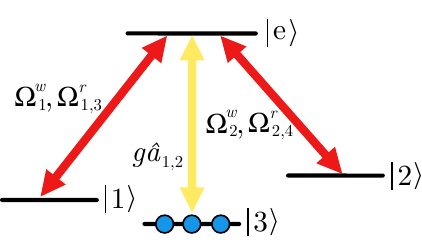}
\caption{The tripod-type atomic configuration. Rabi frequencies of the driving field: $\Om_{1,2}^w$ are at the mapping stage, $\Om_{1,2,3,4}^r$ are at the readout stage. $\a_{1,2}$ are quantum slow amplitudes of the signal field, $g$ is a coupling constant.}
\L{Fig1}
\end{figure}
\begin{enumerate}
\item MP 1. The driving field is switched on with Rabi frequency $\Om_1^w = \Om$ in atomic transition $|1\> - |e\>$, concurrently with the first signal field $\a_1$ in transition $|3\> - |e\>$. Following the complete entry of the signal pulse $\a_1$  into the medium, the driving field is switched off.
\item MP 2. The second driving field is switched on with Rabi frequency $\Om_2^w = \Om$ in atomic transition $|2\> - |e\>$, concurrently with the second signal field $\a_2$ in transition $|3\> - |e\>$. After the pulse $\a_2$ is written, the driving field is switched off.
\item RP 1. Simultaneously, two driving fields are switched on in-phase: one with a Rabi frequency of $\Om_1^r$ in the atomic transition $|1\> - |e\>$ and the second with a Rabi frequency of $\Om_2^r$ in the atomic transition $|2\> - |e\>$. The condition $(\Om_1^r)^2 + (\Om_2^r)^2 = \Om^2$ must be satisfied. Following the readout of the first mixed field $\a_+$, the driving fields are deactivated.
\item RP 2. Simultaneously, two driving fields are switched on in counter-phase: the first is $\Om_3^r$ in the atomic transition $|1\> - |e\>$ and the second is $\Om_4^r$ in $|2\> - |e\>$. Also, the condition $(\Om_3^r)^2 + (\Om_4^r)^2 = \Om^2$ must be satisfied. Consequently, the second mixed field $\a_-$ is read out.
\end{enumerate}

The high-speed quantum memory regime is considered, wherein the times of mapping and readout are much less than the lifetime of atoms in the excited state $|e\>$. This approach enables the Langevin noise operators to be disregarded within the framework of the Heisenberg picture, in accordance with the fluctuation-dissipation theorem \cite{Scully-1997}.

The use of the interaction Hamiltonian in the dipole approximation and the rotating wave approximation results in the derivation of a concise system of equations that delineate the field $\a$ and spin wave $\b$ canonical amplitudes of interest \cite{Losev-2016}
\BEA
&&\partial_z \a_{1,2}(z, t) = -g \sqrt{N} \c(z, t), \L{2}\\
&&\partial_t \c(z, t) = g \sqrt{N} \a_{1,2}(z, t) + \Om \, \b_{1,2}(z, t), \L{3}\\
&&\partial_t \b_{1,2}(z, t) = - \Om \, \c(z, t). \L{4}
\EEA
The constant $g$ defines a coupling in the interaction between weak quantum fields and the medium. Driving fields are bright light. $\b_{1,2}$ and $\c$ are operators of the collective spin waves, which are associated with the coherences of the individual atoms \cite{Losev-2016, Losev-2017}. These operators satisfy the canonical commutation relations $[\b_{1,2}(z,t), \b_{1,2}^\dag(z^\pr,t)] = \delta(z - z^\pr)$, $[\c(z,t), \c^\dag(z^\pr,t)] = \delta(z - z^\pr)$. $N$ is the linear concentration of atoms. 

Solutions to this system for spin wave operators $\b_{1,2}$ are of the following form \cite{Losev-2017, Golubeva-2012}:
\BE
\b_{1,2} (z) = - g \sqrt{N} \Om \! \int^{T \6!}_0 dt^\pr \, \a_{1,2} (T - t^\pr) \: G_{ab}(t^\pr, z) + \v, \L{5}
\EE
here $T$ is the time duration of any mapping or readout stage. The quantum electromagnetic field $\a_{1,2}$ are mapped into spin wave operators $\b_{1,2}$ by means of an integral transformation with a kernel $G_{ab}(t,z)$. Operator $\v$ defines all vacuum components of the system and formally keeps operators $\b_{1,2}$ as canonical variables.

It is not difficult to see that the system of equations at the readout stage, in which two driving fields interact with the medium simultaneously, will take the following form
\BEA
&&\partial_z \a_\pm(z, t) = -g \sqrt{N} \c(z, t), \L{6}\\
&&\partial_t \c(z, t) = g \sqrt{N} \a_\pm(z, t) + \Om \, \b_\pm(z, t), \L{7}\\
&&\partial_t \b_\pm(z, t) = - \Om \, \c(z, t). \L{8}
\EEA
Here
\BE
\b_+ = (\Om_1^r \, \b_1 + \Om_2^r \, \b_2) / \Om, \L{9}
\EE
\BE
\b_- = (\Om_3^r \, \b_1 - \Om_4^r \, \b_2) / \Om, \L{10}
\EE
are also collective canonical operators of the atomic medium \cite{Losev-2016, Losev-2017}.

The system of equations (\ref{6})-(\ref{8}) gives the solution for the retrieved fields
\BE
\a_\pm (t)
= - g \sqrt{N} \Om \! \int^{L \6!}_0 dz \, \b_\pm (z) \, G_{ba}(z, t) + \v, \L{11}
\EE
wherein $L$ is a memory cell length.

In order to express the final form for $\a_\pm$ through the input fields $\a_{1,2}$, it is necessary to substitute (\ref{9}), (\ref{10}) and (\ref{5}) into (\ref{11}) and use the relationship between the kernel of the full memory cycle $G(t,t^\pr)$ (both mapping and readout procedure) and the kernels of corresponding mapping $G_{ab}(z, t)$ and readout stages $G_{ba}(t^\pr, z)$. We consider the case of the backward retrieval. That is, when the signal and driving pulses propagate in the same direction during the mapping stage, and  the driving pulse propagates in the opposite direction during the reading out stage. For this situation, the kernel of the full memory cycle  \cite{Losev-2017}
\BE
G(t, t^\prime) = \frac12 \int^{L \6!}_0 dz \, G_{ab}(z, t) G_{ba}(t^\pr\!, z). \L{12}
\EE

After the dimensionless procedure \cite{Golubeva-2012} using
\BE
z \leftarrow (2 \, g^2 N / \Om) \, z \quad \mbox{and} \quad t \leftarrow \Om \, t, \L{13}
\EE
we get
\BE
\a_+ (t) = \frac1{\Om} \! \int^{T \6!}_0 dt^\pr ( \Om_1^r \a_1 (T - t^\pr)
+ \Om_2^r \a_2 (T - t^\pr) ) G(t, t^\pr) + \v, \L{14}
\EE
\BE
\a_- (t) = \frac1{\Om} \! \int^{T \6!}_0 dt^\pr ( \Om_3^r \a_1 (T - t^\pr)
- \Om_4^r \a_2 (T - t^\pr) ) G(t, t^\pr) + \v. \L{15}
\EE
Both the mapping and readout procedures are performed in two steps. The first signal field $\a_1(z, t)$ is mapped into the first spin wave $\b_1(z)$ during time $T$, the second signal field $\a_2(z, t)$ into the second spin wave $\b_2(z)$ also during time $T$. The first mixed spin wave $\b_+(z)$ is retrieved into the first output field $\a_+(t)$ from the memory cell, and the second mixed spin wave $\b_-(z)$ is retrieved into the second output field $\a_-(t)$.
%
%
%-------------------------------------------
\section{The full memory cycle eigenmodes}\L{III}
The process can be analysed by examining the eigenmodes of the integral operator of the full memory cycle \cite{Tikhonov-2014}:
\BE
\int^{T\6!}_0 dt^\pr G(t, t^\pr) \varphi_i(t^\pr) = \sqrt{\lambda_i} \, \varphi_i(t). \L{16}
\EE
Here $\varphi_i(t)$ are eigenfunctions and $\sqrt{\lambda_i}$ are eigenvalues. The functions form the complete orthonormal set:
\BEA
&&\int_0^{T \6!} dt \,\varphi_i^* (t) \varphi_j(t) = \delta_{ij}, \L{17}\\
&&\sum_i \varphi_i^* (t) \varphi_i (t^\pr) = \delta(t - t^\pr), \L{18}
\EEA
then
\BE
G(t, t^\pr) = \sum_i \! \sqrt{\lambda_i} \, \varphi_i (t) \, \varphi_i^* (t^\pr). \L{19}
\EE
It is also possible to decompose amplitudes $\a(t)$ over the complete set of functions $\varphi_i (t)$: 
\BE
\a(t) = \sum_i \e_i \, \varphi_i (t), \L{20}
\EE
wherein $\e_i$ are discrete canonical operators
\BE
\h{e}_i = \int_0^{T \6!} dt \, \psi_i^\ast(t)\a(t),\quad \[\h{e}_i, \h{e}_j^\dag \] = \delta_{ij} . \L{21}
\EE
Substituting (\ref{19}), (\ref{20}) into (\ref{14}), (\ref{15}), and utilising conditions (16), (17), results in the following:
\BE
\e_{+,i} = \frac{\sqrt{\lambda_i}}{\Om} \( \Om_1^r \e_{1,i} + \Om_2^r \e_{2, i} \) +  \mathcal{N}_i \e_{v,i}, \L{22}
\EE
\BE
\e_{-,i} = \frac{\sqrt{\lambda_i}}{\Om} \( \Om_3^r \e_{1,i} - \Om_4^r \e_{2, i} \) +  \mathcal{N}_i \e_{v,i}. \L{23}
\EE
Here $\mathcal{N}_i = \sqrt{1- \lambda_i}$ is the normalisation constant.
It is possible to combine the obtained equations by means of input-output matrix transformation $\mathcal{M}_i$
\BE
\begin{pmatrix} 
\e_{+,i}\\ 
\e_{-,i}
\end{pmatrix}
=
\mathcal{M}_i
\begin{pmatrix} 
\e_{1,i}\\ 
\e_{2,i}
\end{pmatrix}
+
\mathcal{N}_i
\begin{pmatrix} 
\e_{v,i} \!\\ 
\e_{v,i} \!
\end{pmatrix}, \L{24}
\EE
\BE
\mathcal{M}_i = 
\frac{\sqrt{\lambda_i}}{\Om}
\begin{pmatrix} 
\Om_1^r & \Om_2^r\\ 
\Om_3^r & - \Om_4^r 
\end{pmatrix}. \L{25}
\EE
It is assumed that the transformation $\mathcal{M}_i$ should be unitary ($\Om_4^r = \Om_1^r$, $\Om_3^r = \Om_2^r$). The matrix is parametrized with the angle $\theta_i$, which leads to a form similar to a usual beam splitter \cite{Campos-1989} for fixed phase values
\BE
\mathcal{M}_i
=
\begin{pmatrix} 
\cos \theta_i & \ \sin \theta_i\\ 
\sin \theta_i & - \cos \theta_i
\end{pmatrix}, \L{26}
\EE
wherein
\BEA
&&\cos \theta_i = \frac{\Om_1^r}{\Om} \sqrt{\lambda_i}, \L{27}\\
&&\sin \theta_i = \frac{\Om_2^r}{\Om} \sqrt{\lambda_i}, \L{28}
\EEA
and
\BE
\Om^2 = (\Om_1^w)^2 = (\Om_2^w)^2
= (\Om_1^r)^2 + (\Om_2^r)^2 = (\Om_3^r)^2 + (\Om_4^r)^2. \L{29}
\EE
It can be concluded that mapping in only one mode with eigenvalue $\lambda_i = 1$ provides an exact analogue to a beam splitter with arbitrary values of the splitting ratios:
\BE
\begin{pmatrix} 
\e_+\\ 
\e_-
\end{pmatrix}
=
\begin{pmatrix} 
\cos \theta & \ \sin \theta\\ 
\sin \theta & - \cos \theta
\end{pmatrix}
\begin{pmatrix} 
\e_1\\ 
\e_2
\end{pmatrix}. \L{30}
\EE
%
%
%-------------------------------------------
\section{Conclusion}\L{V}
It has been demonstrated that analysing input field mixing processes in the language of memory eigenmodes $\e_i$ yields the equation (\ref{24}) similar to the equation for the beam splitter with accuracy to vacuum losses $\e_{v,i}$. Operating with only one memory eigenmode, for eigenvalue $\lambda_i$ equal to 1, makes the tripod memory analogous to a beam splitter with arbitrary values of the splitting ratios.
As shown in \cite{Tikhonov-2014}, the selection of defined values for cell length $L$ and mapping time $T$ can reduce the number of eigenmodes in which the mapping and readout operations are most efficient. So in \cite{Tikhonov-2014} the eigenvalues of only the first two modes are close to unity $\lambda_1 = 1$, $\lambda_2 = 0.9$. The eigenvalues of the other modes $\lambda_i$ ($i = 3, 4, ...$) are close to zero. Furthermore, an additional feature of high speed memory is that the first two modes are associated with different time intervals of readout \cite{Tikhonov-2014}. Thus, selecting the necessary experiment parameters can allow working with specific eigenmodes of memory.

The use of quantum memory as a beam splitter has an obvious feature. The mixed fields must reach the memory cell at different times. Concurrently, the emitted fields are also separated in time. These conditions can offer a significant advantage over a usual beam splitter in quantum information protocols.
%
%
%
%-----------------------------------------------------------

%

\begin{thebibliography}{99}
%
\bibitem{Paspalakis-2002}Paspalakis, E.,  and Knight, P. L. 2002, \textit{J. Mod. Opt.} \textbf{49}, 87, doi: 10.1080/09500340110060092
%
\bibitem{Paspalakis-2002PRA}Paspalakis, E., Kylstra, N. J., and Knight, P. L. 2002, \textit{Phys. Rev. A} \textbf{65}, 053808, doi: 10.1103/PhysRevA.65.053808
%
\bibitem{Wang-2004}Wang, T., Kostrun, M., and Yelin, S. F. 2004, \textit{Phys. Rev. A} \textbf{70}, 053822, doi: 10.1103/PhysRevA.70.053822
%
\bibitem{Unanyan-2004}Unanyan, R. G., Pietrzyk, M. E., Shore, B. W., and Bergmann, K. 2004, \textit{Phys. Rev. A} \textbf{70}, 053404, doi: 10.1103/PhysRevA.70.053404
%
\bibitem{Raczynski-2006}Raczynski, A., Rzepecka, M., Zaremba, J., and Zielinska-Kaniasty, S. 2006, \textit{Opt. Comm.} \textbf{260}, 73, doi: 10.1016/j.optcom.2005.10.021
%
\bibitem{Moller-2007}Moller, D., Madsen, L. B., and Molmer, K. 2007, \textit{Phys. Rev. A} \textbf{75}, 062302, doi: 10.1103/PhysRevA.75.062302
%
\bibitem{Yang-2015}Yang, S.-J., Bao, X.-H., and Pan, J.-W. 2015, \textit{Phys. Rev. A} \textbf{91}, 053805, doi: 10.1103/PhysRevA.91.053805
%
\bibitem{Shou-2019PRA99}Shou, C. and Huang, G. 2019, \textit{Phys. Rev. A} \textbf{99}, 043821, doi: 10.1103/PhysRevA.99.043821
%
\bibitem{Shou-2019PRA100}Shou, C., and Huang, G. 2019, \textit{Phys. Rev. A} \textbf{100}, 063844, doi: 10.1103/PhysRevA.100.063844
%
\bibitem{Raczynski-2007}Raczynski, A., Slowik, K., Zaremba, J., and Zielinska-Kaniasty, S. 2007, \textit{Opt. Comm.} \textbf{279}, 324, doi: 10.1016/j.optcom.2007.07.025
%
\bibitem{Raczynski-2007PRA}Raczynski, A., Zaremba, J., and Zielinska-Kaniasty, S. 2007, \textit{Phys. Rev. A} \textbf{75}, 013810, doi: 10.1103/PhysRevA.75.013810
%
\bibitem{Losev-2016}Losev, A. S., Tikhonov, K. S., Golubeva, T. Y., and Golubev, Y. M. 2016, \textit{J. Phys. B: At. Mol. Opt. Phys.} \textbf{49}, 195501, doi: 10.1088/0953-4075/49/19/195501
%
\bibitem{Losev-2017}Losev, A. S., Golubeva, T. Yu., and Golubev, Y. M. 2017, \textit{Laser Phys. Lett.} \textbf{14}, 055208, doi: 10.1088/1612-202X/aa69f8
%
\bibitem{Losev-2020}Losev, A. S., Golubeva, T. Yu., Manukhova, A. D., and Golubev, Yu. M. 2020, \textit{Phys. Rev. A} \textbf{102}, 042603, doi: 10.1103/PhysRevA.102.042603
%
\bibitem{Scully-1997}Scully, M. O., and Zubairy, M. S. 1997, \textit{Quantum Optics} (Cambridge: Cambridge University Press)
%
\bibitem{Golubeva-2012}Golubeva, T., Golubev, Y. M., Mishina, O., Bramati, A., Laurat, J., and Giacobino, E. 2012, \textit{Eur. Phys. J. D} \textbf{66}, 275, doi: 10.1140/epjd/e2012-20723-3
%
\bibitem{Tikhonov-2014}Tikhonov, K., Samburskaya, K., Golubeva, T., and Golubev, Yu. 2014, \textit{Phys. Rev. A} \textbf{89}, 013811, doi: 10.1103/PhysRevA.89.013811
%
\bibitem{Campos-1989}Campos, R. A., Saleh, B. E. A., and Teich, M. C. 1989, \textit{Phys. Rev. A} \textbf{40}, 1371
%
\end{thebibliography}
\end{document}